\documentclass[twocolumn,prl]{revtex4}

\ifx\pdfoutput\undefined
  \usepackage[dvips]{graphicx}
\else
  \usepackage[pdftex]{graphicx}
\fi

\usepackage{dcolumn}
\usepackage{amsmath}
\usepackage{latexsym}

\begin{document}
\title[Short Title]{Radio-frequency Bloch-transistor
Electrometer}

\author{A.~B.~Zorin}
\affiliation{Physikalisch-Technische Bundesanstalt, Bundesallee
100, 38116 Braunschweig, Germany}%

\date{December 12, 2000}

\begin{abstract}
A quantum-limited electrometer based on charge modulation of the
Josephson supercurrent in the Bloch transistor inserted into a
superconducting ring is proposed. As this ring is inductive
coupled to a high-Q resonance tank circuit, the variations of the
charge on the transistor island (input signal) are converted into
variations of amplitude and phase of radio-frequency oscillations
in the tank. These variations are amplified and then detected.
The output noise, the back-action fluctuations and their
cross-correlation are computed. It is shown that our device
enables measurements of the charge with a sensitivity which is
determined by the energy resolution of its amplifier, that can be
reduced down to the standard quantum limit of ${1 \over 2}
\hbar$. On the basis of this setup a "back-action-evading" scheme
of the charge measurements is proposed.

\end{abstract}

\maketitle

\section{Introduction}
The single electron transistor (SET) whose operation is based on
correlated electron tunneling in small-capacitance double
junctions has significantly extended the possibilities of modern
experiments. This remarkable device with sub-electron sensitivity
to the charge induced on its central electrode (island) has made
it possible to study the electron transport and noise processes in
various mesoscopic structures (see, for example, the review by
Likharev \cite{Likh-obz}). In recent years, especially after the
encouraging experiment by Nakamura $et$ $al$. \cite{Nakam}, the
possibility of using SET electrometers for measuring the quantum
state of the charge qubit (Cooper-pair box) has been extensively
discussed \cite{Karlsruhe}. In such measurements both the
sensitivity of the detector (electrometer) to the input signals
and its destructive back-action on the quantum mechanical state of
the box are of prime importance. The detector's figure of merit,
which takes into account the back-action effect, is the energy
resolution in the unit bandwidth $\epsilon$. According to the
quantum mechanical uncertainty principle for a phase-insensitive
detector, the figure $\epsilon \geq \hbar/2$. Its ultimate value
of $\hbar/2$ (the so-called standard quantum limit - SQL) can be
approached by a perfect (quantum-limited) device \cite{amplifier}.

The normal-state metallic SET operating in the regime of
sequential tunneling of electrons drops out of the category of
perfect devices. For the usual case of a high tunneling resistance
of junctions $R_t \gg R_Q$ (here $R_Q \equiv h/4e^2 \approx
6.5$~k$\Omega$ is the resistance quantum), the value $\epsilon \gg
\hbar/2$ \cite{kor1}. One can, in principle, approach SQL
\cite{Aver-c-m, Maassen-c-m} by using SET with $R_t \sim R_Q$ and
operating it in the co-tunneling regime at very low voltage bias.
However, in this case the output signal of the electrometer is
vanishingly small so that the regime can hardly be practical.

In contrast to the SET operating on "normal carriers", i.e.
electrons, its superconducting counterpart with appreciable
strength of Josephson coupling $E_J$ in the junctions, i.e. the
Bloch transistor \cite{Bl-tr}, can operate in the regime of a
gate-controlled supercurrent at zero quasiparticle current. In
this regime the charge carriers are the Cooper pairs with charge
$2e$, and their transfer across the junctions occurs without power
dissipation in the transistor. Reading-out of the critical
current value can be performed by measuring the voltage across the
resistor with $R_s \ll R_Q$, shunting the transistor \cite{Z-PRL}.
Although this electrometer is a quantum-limited device, its
implementation still suffers from low conversion factor
\cite{Lotkh}.

In this paper we propose an electrometer with the Bloch
transistor inserted into a superconducting ring which is
inductively coupled to the radio-frequency-driven resonance tank
circuit. In contrast to the so-called rf-SET electrometer
\cite{Schoel} based on a charge-dependent dissipation in a
resonance circuit containing a normal SET, the Bloch transistor
controls the ac supercurrent in the loop and, hence, the
effective reactance of the tank circuit. As a result, both the
amplitude and phase of oscillations in the tank circuit depend on
the island charge. This mode of electrometer operation is similar
to that of a single-junction superconducting quantum
interferometer device (SQUID) with small critical current
(non-hysteretic regime) \cite{Hansma}. Here we compute the
characteristics of our electrometer and demonstrate its potential
for qubit measurements.

\begin{figure}
\begin{center}
\includegraphics[width=3.2in]{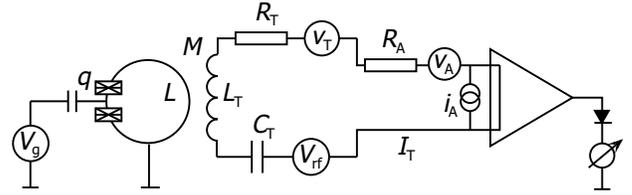}
\caption{Electric diagram of the rf-Bloch-electrometer comprising
the grounded superconducting ring including the Bloch transistor
with capacitively coupled gate, the series-resonance tank circuit
driven by source $V_{\rm rf}$, an amplifier and an amplitude (or
phase) detector.} \label{EqvSchm}
\end{center}
\end{figure}

\section{Model}

The equivalent electric diagram of the electrometer comprising
the sources of fluctuations is presented in Fig.~1. The
characteristic Josephson coupling energies in the first and
second junctions of the transistor are assumed to be not very
different, $E_{J1} \simeq E_{J2}$; so we will characterize both of
them by the parameter $E_J = \Phi_0 I_{c0}/2\pi$, where $\Phi_0 =
h/2e \approx 2.07\times10^{-15}$~Wb is the flux quantum and
$I_{c0}$ is the nominal critical current of individual junctions.

The charge-sensitive element of the device is the transistor
island. We assume that the total capacitance of the island $C =
C_1+C_2+C_g$ (here $C_{1,2}$ and $C_g$ are the capacitances of
the tunnel junctions and the coupling capacitance to the gate,
respectively) is sufficiently small. The corresponding charging
energy $E_c = e^2/2C$, the Josephson energy $E_J$ and the energy
gap $\Delta$ of the superconducting material the transistor made
of should obey the condition:
\begin{equation}
\label{energy}
\Delta > E_c \sim E_J \gg k_BT,
\end{equation}
where $T$ is the temperature. The leftmost inequality ensures
blockade of quasiparticle tunneling across the junctions owing to
the even-odd parity effect on the superconducting island
\cite{Parity}.

The relation chosen, $E_J/E_c \equiv \lambda \sim 1$, first
ensures substantial modulation of the supercurrent in the whole
range of variation of the polarization charge on the island, $-e
\leq q \leq e$ \cite{Z-PRL, Z-IEEE}. Secondly, the width of the
forbidden band in the energy spectrum ($\approx E_J$ at $\lambda
\lesssim 1$) \cite{LikZor} is large enough to prevent thermal
excitation of higher Bloch bands leading to a reduction of the
resultant critical current and of the depth of its modulation by
the gate. For $\lambda \approx 1$ the critical current of the
transistor $I_c(q) = \alpha(q) I_{c0}$, with the value $\alpha$
varied in the range from 0.24 ($q=0$) to 0.56 ($q=\pm e$)
\cite{Z-PRL}. The supercurrent-charge-phase relation is then
approximated by the formula $I_s = I_c(q) \sin \varphi$
\cite{Z-IEEE}.

Due to finite Josephson coupling the effective capacitance of the
electrometer island becomes non-linear \cite{LikZor, Z-PRL}. For
$-{1 \over 2}e \lesssim q\lesssim {1 \over 2}e$ its value is
$C'(q)= \beta(q) C$, where the factor $\beta(q) \gtrsim 1$ can be
assumed to be constant for moderate $\lambda \lesssim 1$.

The inductance $L$ of the superconducting ring incorporating the
Bloch transistor should obey two conditions:
\begin{equation}
\label{L} \ell = 2\pi L I_c(q)/\Phi_0 < 1\quad{\rm and}
\quad\Phi_0^2/2L \gg k_BT.
\end{equation}
The first relation ensures the single-valued dependence of the
total flux $\Phi=\Phi_e-LI_c(q)\sin(2\pi\Phi/\Phi_0)$ threading
the loop on the external flux $\Phi_e = \Phi_{\rm dc} + \Phi_{\rm
rf}$ applied to the loop (see, e.g., Ref.~\cite{KK-book}). The
constant flux $\Phi_{\rm dc}$ can be induced by dc current
through an auxiliary coil (not shown in Fig.~1), while flux
$\Phi_{\rm rf}$ is induced by the tank circuit. For sufficiently
small values of $\ell$, the flux $\Phi\approx\Phi_e$ and the
Josephson phase $\varphi\equiv 2\pi\Phi/\Phi_0 \approx
2\pi\Phi_e/\Phi_0$. The second relation in Eq.~(\ref{L}) ensures
an exponential smallness of thermodynamic fluctuations of flux
$\Phi$. Thus, the Josephson phase $\varphi$ across the transistor
behaves almost as a classical variable whose value and (small)
fluctuations are determined by current in the tank circuit.

The eigenfrequency of the tank circuit $\omega_0 = (L_TC_T)^{-{1
\over 2}}$ and the frequency $\omega \approx \omega_0$ of the rf
drive $V_{\rm rf}=V_{\omega}\cos \omega t$ should be sufficiently
low, i.e. $\omega \ll \omega_J \equiv E_J/\hbar \sim E_c/\hbar$,
and, therefore, not excite the Bloch transistor by means of an
alternative Josephson phase $\varphi(t)$. In our model, $\varphi$
is considered a slowly-varied parameter in the Hamiltonian
\cite{Z-PRL} of the transistor system. The quality factor is $Q =
\omega L/R_\Sigma = (\omega C_TR_\Sigma)^{-1} \gg1$, where
$R_\Sigma = R_T+R_A$ is the total series resistance of the tank
circuit. The dimensionless coupling coefficient is
$\kappa=M/(LL_T)^{1 \over 2} \ll 1$, where $M$ is the mutual
inductance, so that the product
\begin{equation}
\label{k2Ql} \kappa^2 Q \ell > 1. \end{equation} A similar regime
of operation of single-junction (rf) SQUIDs, proposed by Danilov
and Likharev \cite{DanLikh}, offers a significant experimental
advantage in the sense of a large transfer coefficient
\cite{Shnyr}.

The amplifier is characterized by the low active input impedance
$R_A$ and uncorrelated sources of voltage ($v_A$) and current
($i_A$) fluctuations \cite{amplifier}. These two sources and the
source associated with losses in the tank circuit, $v_T$, have
spectral densities
\begin{equation}
\label{S} S_{V,I}(\omega)= {2 \over \pi}\Theta_{A}
R_{A}^{\pm1}\quad {\rm and}\quad S_T(\omega)= {2 \over \pi
}\Theta_T R_T,
\end{equation}
respectively, $\Theta_{A,T}=({\hbar \omega \over 2}) \coth
({\hbar \omega \over 2k_B T_{A,T}})$ with $T_A$ $(T_T)$
symbolizing the noise temperature of the amplifier (temperature
of the tank circuit) in the classical limit $k_BT_{A,T} \gg \hbar
\omega$. Only these three sources of fluctuations are considered
in our model \cite{dissipation}.

\section{Dynamics}

As long as dynamic equations describing the rf-SQUID were solved
elsewhere by the method of harmonic balance
\cite{KK-book,DanLikh,DLS}, we skip mathematical details and
focus chiefly on the results.

Due to large $Q$ and weak coupling $\kappa$, the steady
oscillations of the tank circuit current, $I_T= I_a \cos(\omega t
+ \vartheta)$, and the Josephson phase,
\begin{equation}
\label{fi} \varphi =a \cos(\omega t + \vartheta) +\varphi_0,
\end{equation}
are quasi-harmonic with slowly varying parameters $a=2\pi M
I_a/\Phi_0$ and $\vartheta$ and constant phase
$\varphi_0=2\pi\Phi_{\rm dc}/\Phi_0$. The dependence of the
dimensionless amplitude $a$ on detuning $\xi_0 =
(\omega-\omega_0)/\omega_0$ is shown in Fig.~2. At a sufficiently
large amplitude $V_\omega$ of the driving voltage this dependence
is multi-valued. This property allows high values (theoretically
infinite) of the conversion coefficients "charge-to-amplitude" and
"charge-to-phase" to be realized. Because of the shift of the
resonance frequency in the tank circuit coupled to the
electrometer loop, the effective detuning is $\xi = \xi_0 -
\kappa^2 \ell(q) \cos\varphi_0 J_1(a)/a$. Here, $J_1$ is the
Bessel function of the first order. These peculiar curves are
typical of the rf-SQUID (see, e.g., Ref.~\cite{KK-book}).

\begin{figure}
\begin{center}
\includegraphics[width = 3.4in]{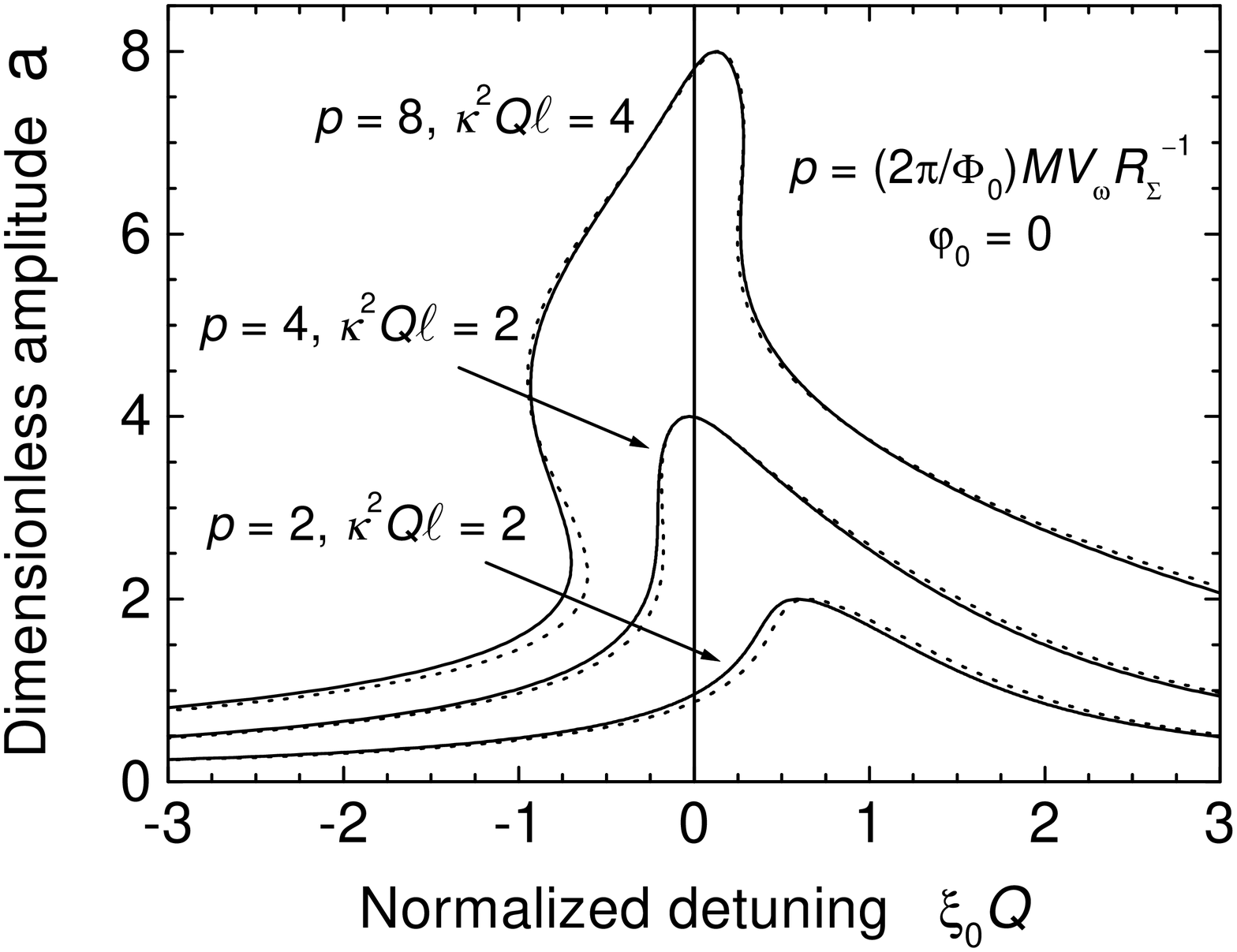}
\caption{Resonance curves of the tank circuit for different
values of the drive amplitude $V_\omega$ and the value of product
$\kappa^2 Q \ell$. The dotted lines correspond to a 10\% increase
in critical current $I_c(q)$ of the transistor.} \label{EqvSchm}
\end{center}
\end{figure}

The coefficients $\eta_a = \left|{\partial I_a \over
\partial q} \right|$ and $\eta_\vartheta =I_a \left|
{\partial \vartheta \over \partial q} \right|$ governing the
transformation of small charge variations $\delta q$ into
variations of two orthogonal components of ac current in the tank,
$\delta I_a$ and $I_a \delta\vartheta$, are expressed as $\eta_a =
|\xi| \eta_0$ and $\eta_{\vartheta} = (2Q)^{-1} \eta_0$,
respectively. Here, the factor
\begin{equation}
\label{H_0} \eta_0 = \mu {M \over L_T}
\left|{\cos\varphi_0\,J_1(a) \over {(2Q)^{-2}+\xi
\tilde{\xi}}}\right|,
\end{equation}
the transfer function $\mu=\left|{\partial I_c \over \partial q}
\right|$ and dynamic detuning $\tilde{\xi} = \xi_0-\kappa^2\ell
\cos\varphi_0 J_1^\prime(a)$. Equation (\ref{H_0}) in particular
shows that zero magnetic flux $\Phi_{\rm dc}$ giving
$\varphi_0=0$ and, hence, $\cos \varphi_0=1$, ensures a maximum of
either $\eta_a$ and $\eta_\vartheta$. This is in contrast to the
rf-SQUID operating ultimately at $|\sin\varphi_0| = 1$, i.e. at
nonzero dc flux \cite{DLS}. The optimum amplitude of the rf-drive
should give the value of $a\approx 1.8$ corresponding to the
maximum value of $(J_1)_{\rm max}=j_1 \approx 0.58$.

\section{Noise figures}

For the amplitude (phase) detection of a low-frequency signal
($\omega_s \ll\omega$), the output resolution in bandwidth $\Delta
f$,
\begin{equation}
\label{dq} \delta q_x = \langle{\tilde{q}}^2\rangle^{1 \over 2} =
\eta_{a,\vartheta}^{-1}\, (2\pi\,S_{a,\vartheta}\,\Delta f)^{1
\over 2},
\end{equation}
is determined by the spectral density of the in-phase
(out-of-phase) fluctuations of the current flowing through the
amplifier,
\begin{equation}
\label{S_I} S_{a,\vartheta} = {g_{a,\vartheta}(\xi, \tilde{\xi})
\over R_\Sigma^2}[S_T(\omega)+S_V(\omega)]+S_I(\omega),
\end{equation}
where $g_a(\xi, \tilde{\xi}) =Q^{-2}(Q^{-2}+4\xi^2)(Q^{-2}+4\xi
\tilde{\xi})^{-2}=g_\vartheta(\tilde{\xi},\xi)$. The {\it output}
noise in the energy representation $\epsilon_I =
\langle{\tilde{q}}^2\rangle /(2C' \Delta f)$ finally takes the
form
\begin{equation} \label{E_I}\epsilon_I^{(a,\vartheta)} = {b d_{a,\vartheta} \over
\kappa^2 Q \ell \,\omega} \left({R_T \over R_\Sigma}\Theta_T+{R_A
\over R_\Sigma}\Theta_A+{R_\Sigma \over g_{a,\vartheta}
R_A}\Theta_A\right).
\end{equation}
Here, the numerical factor $b=\pi I_c/(j_1^2\Phi_0\mu^2 C')$,
while
\begin{equation}
\label{d} d_a = {Q^{-2}+4\xi^2 \over 4\xi^2}\quad {\rm and}\quad
d_{\vartheta} = {Q^{-2}+4\tilde{\xi}^2 \over Q^{-2}}
\end{equation}
for the case of amplitude and phase detection, respectively. Note
that, owing to the large value of product $\kappa^2Q\ell$
(Eq.~(\ref{k2Ql})), the output noise figures
$\epsilon_I^{(a,\vartheta)}$ (they do not include the back-action
effect!) can be made smaller than $\hbar/2$ in the limit $T_T, T_A
\ll \hbar\omega/k_B$.

The electrometer back-action on the source of the input charge is
determined by low-frequency ($\sim\omega_s$) fluctuations of the
electric potential of the transistor island $\tilde{U} = {\Phi_0
\over 2\pi}\mu\, \widehat{\sin\varphi\,\tilde{\varphi}}$
\cite{Z-PRL}. Here $\tilde{\varphi}$ are fluctuations of the
Josephson phase Eq.~(\ref{fi}) and $\widehat{...}$ denotes
averaging over time $\tau$: $2\pi/\omega\ll \tau\ll
2\pi/\omega_s$. Finally, the {\it input} noise figure $\epsilon_U
= C'\langle{\tilde{U}}^2\rangle /(2\Delta f)$ for either regime
is given by
\begin{equation}
\label{E_U}\epsilon_U^{(a)} =\epsilon_U^{(\vartheta)}= {g_{a}
\,\kappa^2 Q \ell \over b\,\omega} \left({R_T \over R_\Sigma}
\Theta_T+{R_A \over R_\Sigma}\Theta_A\right).
\end{equation}

From Eq.~(\ref{fi}) it follows that fluctuations $\tilde{U}$ are
proportional to fluctuations of amplitude $\tilde{a}$; therefore,
these two signals are completely correlated. Due to this fact the
cross-correlation
$\epsilon_{IU}=|\langle{\tilde{q}\tilde{U}}\rangle|/2\Delta f$ in
the regime of amplitude detection has the largest magnitude which
is equal to the geometric mean of $\epsilon_U^{(a)}$
Eq.~(\ref{E_U}) and $\epsilon_I^{(a)}$ with the third term omitted
in Eq.~(\ref{E_I}). Then the energy resolution of a narrow-band
signal \cite{Z-PRL,DLS,Danilov}
\begin{equation}
\label{E-def}\epsilon = \left[\epsilon_I \epsilon_U -
\epsilon_{IU}^2\right]^{1 \over 2}
\end{equation}
is equal to
\begin{equation}
\label{E-obt} \epsilon = \omega^{-1} {\{d_a \Theta_A
[(R_T/R_A)\Theta_T+\Theta_A] \}}^{1 \over 2}.
\end{equation}

This equation shows that the electrometer figure of merit
$\epsilon$ depends crucially on the amplifier characteristic
$\Theta_A$. In particular, for $R_T \Theta_T \ll R_A \Theta_A$ and
detuning $|\xi| \gg (2Q)^{-1}$, the figure
$\epsilon=\Theta_A/\omega$, and its value approaches the SQL of
$\hbar/2$ at $k_B T_A < \hbar\omega$.

\section{Discussion}

We arrive at the remarkable property of the rf-Bloch-electrometer:
it converts an input charge into an output signal introducing only
insignificant noise on the stage preceding the amplifier. This is
because the device operates as a parametric converter $\omega_s
\rightarrow (\omega\pm\omega_s) \rightarrow \omega_s$ (similar to
the single-junction SQUID, see, e.g., Ref.~\cite{KK-book}). In
such a scheme of electrometer (in contrast to other SET
electrometers) the amplifier can be optimized as a {\it separate}
block. In the frequency range of $100-500$~MHz, the
state-of-the-art narrow-band dc-SQUID-based amplifiers make it
possible to almost approach the SQL \cite{Mueck}. For such an
amplifier the required impedance matching with the tank can be
carried out by means of a transformer.

The set of experimental parameters for the Al transistor can be
chosen as follows: $E_J \sim E_c \sim 200~\mu$eV (corresponds to
$C \sim C'/2 \sim 0.2$~fF, $I_{c}\sim 30~$nA and $\omega_J/2\pi
\sim 50$~GHz $\gg \omega/2\pi \sim 300$~MHz), $L \sim 10$~nH
(gives $\ell \sim 0.3$ and $\Phi_0^2/2k_BL \sim 10$~K $\gg T \sim
20$~mK), $Q \sim 300$ (bandwidth $\sim \omega/Q \sim 1$~MHz) and
$\kappa^2 \sim 0.3$. These parameters yield the value $\kappa^2 Q
\ell \sim 30$ that ensures a large transfer coefficient. For the
quantum-limited amplifier the charge resolution is expected to be
equal to $(C'\hbar)^{1 \over 2} \approx 2\times10^{-7}~e/{\rm
Hz}^{1 \over 2}$.

Another important conclusion can be drawn regarding a possible
"back-action-evading" measurement by the rf-Bloch-electrometer.
Such measurement assumes that one quadrature component of internal
noise is "squeezed" to less than SQL \cite{Takahashi}. One of the
ways to do so is to apply to the tank two driving signals with
frequencies $\omega_1$ and $\omega_2$ which obey the relation
$\omega_s=\omega_1-\omega_2$ (see similar proposal for rf-SQUID
in Ref.~\cite{Danilov}). In this "degenerate" mode of operation
\cite{Panov} the device is sensitive to a quadrature component,
say $\hat{X_1}$, of the input ac signal $q =
(\hat{X_1}+i\hat{X_2})e^{i\omega_s t}$ whose Heisenberg
uncertainties obey the relation
$\delta\hat{X_1}\times\delta\hat{X_2} \geq C'\hbar$. As a result,
one side ($\delta\hat{X_1}$) of the "error box" is squeezed while
another ($\delta\hat{X_2}$) is increased, with their product kept
constant.

Finally, there is yet another advantage of the
rf-Bloch-electrometer for qubit measurements: Its transducer (the
ring with transistor) is generically superconducting, the tank
circuit should preferably also be made of superconducting
material. This device when positioned near qubit is, therefore,
free from the normal-electron excitations which may significantly
shorten a decoherence time of qubit.

The experimental work on the radio-frequency Bloch-transistor
electrometer has been started at PTB.

\section{acknowledgments}

The author wishes to thank M.~G\"otz, E.~Il'ichev, V.~V.~Khanin,
A.~Maassen van den Brink, J.~Niemeyer and S.~S.~Tinchev for
stimulating discussions. This work is partially supported by DFG
Grant NI 253/4-1.

\end{document}